\documentstyle[preprint,prb,aps]{revtex}
\begin{document}
\draft

\title{Two-electron state in a disordered 2D island: pairing  caused by the
Coulomb repulsion}

\author{M. E. Raikh$^{(1)}$, L. I. Glazman$^{(2)}$, and L. E. Zhukov$^{(1)}$ }
\address{
$^{(1)}$Physics Department, University of Utah, Salt Lake City, Utah
84112\\
$^{(2)}$Theoretical Physics Institute and Department of Physics, University of
Minnesota, Minneapolis,  Minnesota 55455}

\maketitle
\begin{abstract}

We show the existence of bound two-electron states in an almost depleted
two-dimensional island. These two-electron states are carried by
special compact configurations of four single-electron levels. The existence of
these
states does not require phonon mediation, and is facilitated by the
disorder-induced potential relief and by the electron-electron {\em
repulsion\/}
only. The density of two-electron states is estimated and their
 evolution with  the magnetic field  is discussed.
\end{abstract}
\narrowtext
\tighten

In a  recent experiment\cite{ash1,ash2} the tunneling of
electrons from a semimetallic electrode into the localized states (LS's)
 in a quantum well was studied. By using the electron beam lithography, a dot
with a diameter as asmall as 1 $\mu$m was formed.
 The small size of the dot enabled the authors
 to detect the individual tunneling acts. These acts
 manifested themselves as narrow peaks in the differential capacitance
measured as a function of the  bias applied across the structure.
To introduce the LS's in the $GaAs$ quantum well the neighboring $AlGaAs$
region was doped with $Si$ donors.

 The authors identified the origin of LS's by studying the evolution of
the peak positions with magnetic field, $B$,  perpendicular to
the well.
They assocaited the peaks with electronic states of essentially two types:
 ($\em i$) ground state in the cylindrically symmetric parabolic
potential; the $B$-dependence of the corresponding peaks above certain
$B$ approached the one for the lowest Landau level. Such a parabolic
confinement presumably results from the fluctuations in the
concentration of donors in the barrier.
 ($\em ii$) bound state of an electron at a
$Si$ donor which could migrate into the well during the growth. For
these states the increase of the energy with $B$ was much slower than for
the group ($\em i$).

There is a puzzling feature in the data reported in \cite{ash1}:
 few of the peaks observed were twice as high as isolated
 one-electron
peaks. This suggests that two electrons tunnel into the well at the
same voltage applied. By tracing the $B$-dependencies of these peaks
the authors have ruled out the possibility that they reflect
accidental degeneracies in the energy positions of LS's in some
distant minima. All double peaks retained their height within a
certain range of $B$ and then split into doublets at some critical
value of the magnetic field. Such a behavior indicates that both LS's
involved ``feel'' each other and, thus, are located close in space.
On the other hand it is apparent that two close potential minima
cannot accomodate two electrons at the same bias. Even if the energy
levels are degenerate,
 tunneling of one electron would elevate the
level for the second electron, so that the
subsequent tunneling will occur  at the  bias
larger by the energy of the Coulomb repulsion. The authors mentioned
that the physical mechanism which could resolve this paradox is the
polaronic effect: this effect favors double occupation of LS's in
glassy materials \cite{and75}. Their conjecture was further developed
in \cite{phi95}, where the two-electron state in a hybrid hydrogenic-parabolic
potential in the presence of electron-phonon interaction was considered. The
pair-binding condition used in \cite{phi95} implicitly assumed that the two
electrons share the same lattice deformation, which leads to the enhancement of
the polaronic effect. In fact, the used in~\cite{phi95} polaronic shift  per
electron in the paired state is twice the shift for a single localized
electron. Under this assumption bound two-electron states were found even at
weak electron-phonon interaction, provided that the distance between the
hydrogen-like impurity and the center of the parabolic potential is
larger than $8a_0$, where $a_0$ is the radius of the
hydrogenic state. For such distances the enhanced polaronic shift
overweighs the Coulomb repulsion. However
we find the underlying assumption hard to justify. Indeed, the spatial scale of
the polaronic deformation coincides with the size of a single-electron
state \cite{rash84}; two distant electrons do not share the same deformation,
and, therefore,
the corresponding enhancement of the polaron shift is suppressed.

In the present paper we demonstrate that the double peaks observed
in \cite{ash1} can be naturally explained without invoking the
electron-phonon interactions. Our explanation is based exclusively
on electrostatics. We assume that the electrons are strongly localized and
neglect the overlap of their wave functions
and, correspondingly, the exchange interaction.
 On the other hand the modification of the
Coulomb interaction between  localized electrons due to the
presence of an electrode plays an important role in our picture.
 We show that, in contrast to \cite{phi95}, the
objects responsible for two-electron tunneling are the compact
groups of LS's and calculate the relative portion of the double
peaks.

First of all, let us establish the general criterion for two-electron
tunneling. Consider a cluster of $N$ LS's occupied by $n$ electrons.
The distribution of electrons over LS's corresponds to the {\em  minimal}
possible energy which we denote as $E_N^n$. The position of the
Fermi level in the electrode, $E_F^1$, at which an additional electron
will enter the cluster is determined from the condition $E_F^1 + E_N^n
= E_N^{n+1}$. If  two electrons  enter the cluster, the
corresponding position of the Fermi level, $E_F^2$, satisfies the
relation $2E_F^2 + E_N^n = E_N^{n+2}$. Double peak occurs if $E_F^2 <
E_F^1$. This leads us to the following criterion:
\begin{equation}
\label{1}
E_N^{n+2}+ E_N^n < 2E_N^{n+1}.
\end{equation}
Obviously  condition (\ref{1}) cannot be satisfied if $N=2$.
Formally, if we denote the energies of two LS's as $\varepsilon_1$ and
$\varepsilon_2$ so that $\varepsilon_1< \varepsilon_2$, then $E_2^1 =
\varepsilon_1$,
 $E_2^2 = \varepsilon_1 + \varepsilon_2 + V_{12}$, $V_{12}$ being the
interaction energy  of two electrons occupying the first and the
second LS's. We  see that $E_2^2>2E_2^1$--the inequality opposite to
(\ref{1}).

  Now we will prove that for $N=3$ the occurrence of a double peak
is also forbidden by Eq. (\ref{1}). By analogy to the
consideration above, the  case $n=0$ is obvious for any $N$ since
a single electron on a cluster will occupy the LS with the
 lowest energy level, so
that the condition (\ref{1}) is violated even without the Coulomb repulsion.
Thus the only case to be considered is $n=1$. Let us again order
the energies of LS's: $\varepsilon_1<\varepsilon_2<\varepsilon_3$. Then
$E_3^1=\varepsilon_1$, and $E_3^3=\varepsilon_1+\varepsilon_2+\varepsilon_3 +
V_{12}+ V_{13}+V_{23}$. Now there are three candidates for
$E_3^2$ in accordance with  three variants of occupation of $N=3$
cluster by two electrons. Important is that $E_3^2$ is the {\em
minimal} of these three energies. This means that if the double peak is
possible,
the condition (\ref{1}) should be met when we substitute for $E_3^2$
{\em each} of these energies. Let us choose two of the three candidates
for $E_3^2$, namely: $\varepsilon_1 + \varepsilon_2 + V_{12}$ and
$\varepsilon_1 + \varepsilon_3 + V_{13}$, which correspond to the occupation
of the first and second, and the first and the third LS's respectively.
Substituting them into the right-hand side of Eq. (\ref{1}), we get the
following system of inequalities:
\begin{equation}
\label{2}
\varepsilon_3-\varepsilon_2 < V_{12}-V_{13}-V_{23},
\end{equation}
\begin{equation}
\label{3}
\varepsilon_2-\varepsilon_3 < V_{13}-V_{12}-V_{23}.
\end{equation}
We see that since $V_{23}>0$ the conditions (\ref{2}) and (\ref{3}) are
inconsistent and, hence, the clusters of three LS's cannot provide
double peaks.

Let us turn to the case $N=4$. Because the number of variants increases
dramatically in this case we will restrict our search. Namely we will
assume that the first LS with the lowest enery $\varepsilon_1$ is located
in the center of an equilateral triangle while the other three LS's  with
energies $\varepsilon_2<\varepsilon_3<\varepsilon_4$ are  located in the
vertexes. Then
the energy of repulsion takes only two values,
$V_1$ and $V_2$ (see Fig. 1). The first electron enters the system at
$E_F^1=E_4^1\equiv\varepsilon_1$. It is easy to see that there are only two
energies competing for $E_4^2$, which are: $\varepsilon_1 +\varepsilon_2 + V_1$
and $\varepsilon_2 +\varepsilon_3 + V_2$. All the other two-electron states
have higher energies. Similarly we conclude that there are only two
candidates for $E_4^3$. They are: $\varepsilon_1 + \varepsilon_2 +\varepsilon_3
+
2V_{1} + V_{2}$ and $\varepsilon_2 + \varepsilon_3 + \varepsilon_4 + 3V_2$.
Let us again assume $n=1$
in the condition (\ref{1}). Then, according  to the general procedure, one
should choose the lowest of two
values for $E_4^3$ and check  condition (\ref{1}) with both candidates
for $E_4^2$. If we pick the first candidate for $E_4^3$ and  $\varepsilon_1 +
\varepsilon_2 + V_1$ for $E_4^2$,  this condition reduces to:
 $\varepsilon_3 - \varepsilon_2 < - V_2$, which contradicts our assumption that
the energies are ordered. Thus, the only  remaining option is that
from two candidates for $E_4^3$ the second one has the
lower energy. The corresponding condition for this can be written as
\begin{equation}
\label{4}
\varepsilon_4 - \varepsilon_1 < 2(V_1-V_2).
\end{equation}
Now with $E_4^3 = \varepsilon_2 + \varepsilon_3 + \varepsilon_4 + 3V_2$
 the system of inequalities, resulting from (\ref{1}) in a similar way as
(\ref{2}), (\ref{3}), takes the form
\begin{equation}
\label{5}
(\varepsilon_4-\varepsilon_2)+(\varepsilon_3-\varepsilon_1)< 2V_1-3V_2,
\end{equation}
\begin{equation}
\label{6}
(\varepsilon_4 - \varepsilon_2)-(\varepsilon_3-\varepsilon_1)< - V_2.
\end{equation}
Upon summation of these two inequalities we get
\begin{equation}
\label{7}
\varepsilon_4 - \varepsilon_2 < V_1 - 2V_2.
\end{equation}
On the other hand we have assumed that $\varepsilon_4 > \varepsilon_2$.
Then the principal requirement for a double peak to occur reduces
to: $ V_1 > 2V_2 $.
If this requirement is met, the inequalities (\ref{5}) and (\ref{6}) are
consistent. This is illustrated in Fig. 1 where the graphical solution of the
system (\ref{5}), (\ref{6}) is shown. In principle, one should also check
that  the energies of LS's, satisfying the
system, satisfy the condition (\ref{4}) as well. Note, however, that the
condition (\ref{5}) is stronger than (\ref{4}), so that the latter is
automatically obeyed,
 as can  readily be seen from the following chain of relations:
\begin{equation}
\label{8}
\varepsilon_4-\varepsilon_1<(\varepsilon_4
-\varepsilon_2)+(\varepsilon_3-\varepsilon_1)
<2V_1 - 3V_2 < 2(V_1 - V_2).
\end{equation}

Thus we have demonstrated that, within the restricted geometry considered,
 double peaks can occur provided that $V_1>2V_2$. Obviously, the
relation between $V_1$ and $V_2$ is opposite
if the interaction between the localized electrons is simply the
 Coulomb repulsion. Since  the distance between the first and the second LS
is $\sqrt 3$ times smaller than the distance between the second and the third
LS (see Fig. 1) we have $V_1 = \sqrt 3 V_2$. The situation changes if a
metallic electrode
is placed at a distance $d$ from the plane of the localized electrons.
 Then the Coulomb interaction is modified to
\begin{equation}
\label{9}
V(r)=\frac{e^2}{\kappa}\left(\frac {1}{r} - \frac {1}{\sqrt {r^2 +
4d^2}}\right).
\end{equation}
The modified interaction falls off as $1/r^3$ and we indeed have  $V_1>2V_2$
as soon as the distance between the first and the second LS's exceeds $0.33d$.

If the system (\ref{5}), (\ref{6}) is satisfied the evolution of the
occupation of the cluster with increasing the gate voltage (Fermi level postion
 $E_F$) would be as follows. For $E_F<\varepsilon_1$ all four LS's are empty.
At $E_F=\varepsilon_1$ the first LS in the center of the triangle gets
occupied. As $E_F$ reaches the value $E_F = (\varepsilon_2 + \varepsilon_3 +
\varepsilon_4 + 3V_2 -\varepsilon_1)/2$ an electron from the center moves
to one of the vertexes and two electrons  arrive from the electrode  and
 occupy two other vertexes. Finally at $E_F= \varepsilon_1 + 3V_1$ the LS
in the center gets occupied again.

After realizing that double peaks are possible in principle, we turn to the
question: how frequent are they? One could argue that  double  peaks
are allowed only for extremely rare configurations that do not really
occur in a finite-size sample.  To answer this question we calculate  the
probability that in a cluster of 4 LS's the energies and  distances between
LS's  are arranged in such a way that the two-electron tunneling becomes
possible after the cluster is singly occupied. We start from the
observation that the previous consideration for restricted geometry
becomes general if, instead of (\ref{9}), we assume the model ``hard core''
interaction between the LS's: $V(r)=U$ for $r<d$ and $V(r)=0$ for $r>d$. It
is important that the interaction takes only two values: $U$ and zero. Then
the above analysis for the equilateral triangle applies if the distances,
$r_{ij}$, between the LS's satisfy the following requirements: $r_{12},
r_{13}, r_{14}< d$ and $r_{23}, r_{34}, r_{24}> d$. If these requirements are
met, the conditions for the double peak formation are given by
Eqs.(\ref{5}), (\ref{6}) with $V_1= U$ and $V_2=0$.  It can be verified
directly that for all other configurations of 4 LS's double peaks are
forbidden. Then the calculation of the probability, ${\cal P}$, of the
occurrence
of a double peak  can be performed in a following way: we fix the  position
and the energy of the first LS and find the allowed phase volume for other
three LS's. The advantage of the ``hard core'' interaction is that the
intergrations over coordinates and energies are decoupled from each
other. If we denote with $g$ the density of LS's, then the expression for
${\cal P}$ can be presented as ${\cal P}=g^3 I_1 I_2$, where $I_1$ and $I_2$
are the phase volumes in the energy and coordinate spaces respectively:
\begin{equation}
\label{10}
I_1=\int_{\varepsilon_1}^{\infty}d\varepsilon_2\int_{\varepsilon_2}^{\infty}d\varepsilon_3
\int_{\varepsilon_3}^{\infty}d\varepsilon_4 \theta
(2U+\varepsilon_1+\varepsilon_2-
\varepsilon_3-\varepsilon_4)\theta(\varepsilon_2
+\varepsilon_3-\varepsilon_1-\varepsilon_4),
\end{equation}
\begin{equation}
\label{11}
I_2=\int d{\bf r}_2 \int d{\bf r}_3 \int d{\bf r}_4 \theta(r_{23}-d)
\theta(r_{24}-d)\theta(r_{34}-d)\theta(d-r_{12})\theta(d-r_{13})\theta(d-r_{14}).
\end{equation}

The analytical evaluation of the first integral results in $I_1=U^3/3$.
The integral $I_2$ is obviously proportional to $d^6$; the numerical factor
was found using the Monte-Carlo procedure. Finally we obtain ${\cal P}=
0.611(gUd^2)^3$. We see that the portion of double peaks is governed by
the dimensionless parameter $gUd^2$ which is the ratio of the interaction
energy and the mean level spacing within the size of the ``core.''

It is apparent that for the realistic interaction (\ref{9}) the estimate for
 ${\cal P}$ emerges if one substitutes for $U$ the value $e^2/\kappa d$ ---
 the Coulomb interaction at distance $d$. This gives ${\cal P}\sim
(ge^2d/\kappa)^3$. To find the numerical coefficient, the Monte-Carlo
integration over the  9-dimensional space (6 coordinates and 3 energies) was
performed using the program published in the book \cite{receipt}. The
program generated a random set of dimensionless
(in the units of $d$ and $e^2/\kappa d$)  coordinates and energies,
calculated the values $E_4^1, E_4^2,
 E_4^3$ for the interaction (\ref{9}), and then checked  condition
 (\ref{1}). The numerical factor obtained is $(5.1\pm 0.1)\cdot 10^{-2}$.

The calculation of $g$ poses  a separate problem.
One approach to estimate $g$ is to assume that the random potential
just smears the edge of the  band density of states $g_0=m/\pi\hbar^2$. Then
for
energies not very deep in the tail, $g$ is still of the order of $g_0$.
The product $g_0e^2d/\kappa$ can be rewritten as $d/\pi a_0$, where $a_0=
\hbar^2\kappa/me^2$ is the effective Bohr radius. It may
seem that, if $d$
is large enough, this product could be much larger than 1 .
However this is not the case, since with increasing $d$ the interaction
of the occupied LS's  becomes important. This
leads to  the suppression of the density of states in the
vicinity of the Fermi level (Coulomb gap \cite{ef}, \cite{pol}). For
the interaction (\ref{9}) the energy dependence of $g$
 was studied both analytically
\cite{mog} and by computer simulations \cite{pik}. It was shown that
$g(\varepsilon)= 0.085(d/\kappa e^2)+ \frac{2}{\pi}
(\kappa^2|\varepsilon - E_F|/e^4)$. In our problem the relevant energy scale is
$|\varepsilon - E_F|\sim e^2/\kappa d$ so that $ge^2d/\kappa$ is of the order
of
 1. In view of the ambiguity in $g$, our calculation can be considered only  as
an
estimate showing that ${\cal P}$ is not small. Indeed, the only small
parameter in ${\cal P}$ is the numerical factor $0.051$.  This
factor is, in fact, surprisingly large taking into account that it emerged as
a result of the 9-fold integration.

We have addressed only the energy aspects of the double peak formation. There
is
also a question about the dynamics of the process. Suppose that the occupation
of a single-electron state requires a time $\tau$ ($\tau$ is  inverse
proportional to the probability of tunneling). For a two-electron state with a
binding energy $W$ it can be shown that this time increases  dramatically and
becomes of the order of $\tau^2 W/\hbar$. The reason for the enhancement is
that
two electrons cannot tunnel sequentially because of the energy restrictions.
However, if the temperature $T$ is finite the sequential tunneling becomes
possible due to the smearing of the Fermi distribution in the electrode. Assume
for simplicity that both single-electron energies in a two-electron state are
the same. Then the time required for the sequential occupation of the
two-electron state  is equal to  $\tau \exp(W/2T)$. This time is shorter than
$\tau^2 W/\hbar$ if $T>W/[2 \ln(W\tau/\hbar)]$.

Finally let us discuss the magnetic field dependence of the double peaks.
Note that  conditions (\ref{5}),(\ref{6}) (formation
of a double peak) require  the energies of all four LS's in the cluster
to  be rather close (roughly speaking, they should lie within
the interval of the order of $e^2/\kappa d$). With increasing $B$ each LS
moves up in energy. Important is that the rate of this motion is different
for different LS's. This is obvious if some of LS's originate from donors,
located in the well, while others represent the size quantization levels in
the lateral  fluctuations of the random potential \cite{ash1}, \cite{phi95}.
 Since the rate  for donors
is much  slower,  condition (\ref{1}),  met at $B=0$, will get
violated at  some critical $B$ due to the spread in the level positions. For
higher $B$ the double peak will split into two. Even in the case when all
four components of the double-peak-cluster at $B=0$ are the ground
states in parabolic confinements, their energies will depart from each
other with increasing $B$, thus  causing a splitting of the peak.
 If a confinement is characterized by the position of minimum $U_0$ and
the frequency of the  zero-point motion $\omega_0$, the behavior of the
energy level with $B$ is given by
$\varepsilon(B)=U_0+\hbar(\omega_0^2+\omega_c^2/4)^{1/2}$
($\omega_c$ stands for the cyclotron frequency). Suppose that at $B=0$
two levels, $\varepsilon_1(0)$ and $\varepsilon_2(0)$, are anomalously close in
energy (in order to participate in the cluster). This means that the sum
$U_0^{(1)} + \hbar\omega_0^{(1)}=\varepsilon_1(0)$ is close to $U_0^{(2)} +
\hbar\omega_0^{(2)}=\varepsilon_2(0)$, while separately $U_0^{(1)}$ and
$U_0^{(2)}$ can differ, say, by a factor of two. But in a  strong magnetic
field we have
$\varepsilon_1(B)-\varepsilon_2(B) = U_0^{(1)}-U_0^{(2)}$, so that  departure
is
 the typical fate of the initially aligned levels.

In conclusion, we have demonstrated that  double peaks in the
differential capacitance may result from the interaction-induced
correlations in the occupation numbers of LS's within a cluster. More
conventional consequence  of
these  correlations is that  adding  of $\em one$ electron to the cluster might
cause
a redistribution of neighboring electrons over LS's in order to reduce the
total energy. This  process is similar to the formation of a polaron
by a lattice surrounding an LS.  Note that such a purely electronic
``polaron'' was studied  intensively  by  Efros and Shklovskii and by Pollak
and Ortu\~{n}o (see e.g. the reviews \cite{ef},\cite{pol}) in connection with
the density of states and inelastic transport in the Coulomb glass.
Making a  link to these works, our main result can be reformulated as
follows: for interactions which fall off steeply enough with distance, the
formation of an ``electronic {\em  bipolaron}''  in certain compact clusters
of LS's  is energetically favorable.

The authors are grateful to R.C.~ Ashoori for a valuable comment. One of the
authors (L.~G.) is grateful to E.I.~Rashba for
illuminating discussion on the deformation fields of a polaron. The work at the
University of Minnesota was supported by NSF Grant DMR-9423244.

\begin{figure}
\caption{Cluster of 4 LS's providing a double peak (a).
 Graphical solution of the system (5), (6). Dashed is the region within
which this system is satisfied (b).}
\label{fig1}
\end{figure}
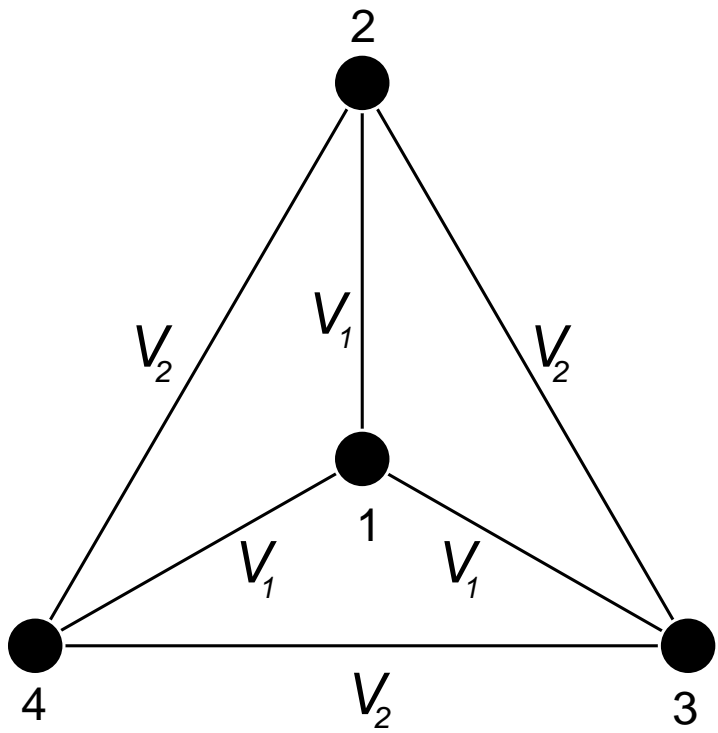
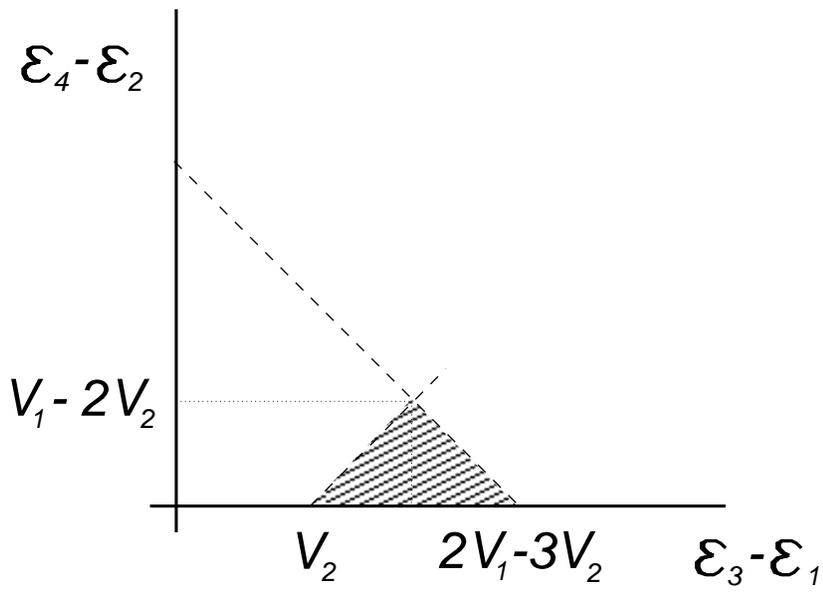
\end{document}